\documentclass[12pt, preprint]{aastex}

\usepackage{amssymb}
\usepackage{amsmath}
\usepackage{epsfig}
\usepackage{graphicx}
\usepackage{natbib}
\usepackage{multirow}
\usepackage{rotating}
\usepackage{longtable}
\usepackage{verbatim}
\usepackage{url}
\usepackage{hyperref}
\usepackage{caption2}

\begin{document}

\title{
Variational Principle for Planetary Interiors
}
\author{Li Zeng\altaffilmark{1,a} and Stein B. Jacobsen\altaffilmark{1,b}
}
\affil{$^1$Department of Earth and Planetary Sciences, Harvard University, Cambridge, MA 02138}

\email{$^a$astrozeng@gmail.com} 
\email{$^b$jacobsen@neodymium.harvard.edu}

\begin{abstract}

In the past few years, the number of confirmed planets has grown above 2000. It is clear that they represent a diversity of structures not seen in our own solar system. In addition to very detailed interior modeling, it is valuable to have a simple analytical framework for describing planetary structures. 
Variational principle is a fundamental principle in physics, entailing that a physical system follows the trajectory which minimizes its action. It is alternative to the differential equation formulation of a physical system. 
Applying this principle to planetary interior can beautifully summarize the set of differential equations into one, which provides us some insight into the problem. From it, a universal mass-radius relation, an estimate of error propagation from equation of state to mass-radius relation, and a form of virial theorem applicable to planetary interiors are derived. 

\end{abstract}
\keywords{planetary interior, action, internal energy}

\section{Introduction}

Variational principle as a fundamental principle bears many applications in mathematics and physics. 
In classical mechanics, treating time as the independent variable, one could describe the motion of a physical system by Newton's Second Law, which boils down to solving a set of coupled differential equations. 
However, in the 18th and 19th-century, an alternative approach was developed based on defining an action for the system as an integral from the initial state to the final state of variable time. The minimization of this action gives the unique evolutionary trajectory of the system in space-time. And it could be easily transformed into the differential equation point of view as the Euler-Lagrange Equation, which is usually a 2nd-order differential equation, or equivalently, the Hamilton Canonical Equations, which are a pair of 1st-order symplectic differential equations. 

Here in this paper, we adopt this idea and apply it to the interior of planets. Instead of treating time as the independent variable as in mechanics, here we treat mass $m$, which is the mass enclosed within radius $r$ as the independent variable. And the volume $\omega$ enclosed in $r$ is taken as the dependent variable where spherical symmetry is assumed. The planet is also assumed to be in a stationary state which evolves slowly so that at every instant its interior is in detail balance. 

Then we derive the action, and the equivalence of the Euler-Lagrange Equation, for the planetary interior. Applying this equation to various Equations of State (EOS) gives us interesting and useful results. Some of the results repeat the results of people's previous works, such as those of polytrope EOS, but in a simpler and neater way, and some of the results are new, such as a universal mass-radius relation for a two-layer rocky planet, and a form of virial theorem applicable to planetary interiors. 

In particular, an emphasis is placed on the power-law EOS, which is equivalent of the polytropic EOS used to derive Lane-Emden Equations in astrophysics. The polytropes were important in developing the early theories of stellar interior structures in the early 20th century~\citep{Eddington:1926, Chandrasekhar:1939, Cox:1968}, as back then, a large quantity of stars were observed but with limited measurement accuracies. Many important results and scaling relations were obtained by applying the polytropes to the ensemble of stars. The situation is now similar as many exoplanets are observed, but with limited accuracies. Thus, the polytrope approach, and modification of which, shall remain valuable when applied to the ensemble of exoplanets, in order to understand the classifications and general properties of them.

\section{Deriving a General Equation of Planetary Interior}

In classical mechanics, the independent variable is time t, and the dependent variable is the coordinate in space such as $x$. The first-order time derivative of $x$ is denoted as $\dot{x}$ (velocity). Lagrangian L=T($\dot{x}$)-V($x$), where the kinetic energy T is a function of $\dot{x}$, and the potential energy V is a function of $x$. 

For planetary interior, the independent variable is mass $m$ and the dependent variable is volume $\omega$ ($=\frac{4\pi}{3}r^3$) chosen for the sake of simplicity. The first-order derivative of $\omega$ with respect to $m$ is denoted as $\dot{\omega}=\frac{d\omega}{dm}=$ specific volume $v=\frac{1}{\rho}$. Then, the question reduces to finding the appropriate action which can describe the system. The key here is to realize that the total action corresponds to the (negative) total energy of the system (action generally has the dimension of energy multiplied by time, however, since here we are considering stationary system, the time-part can be taken out, and the variational principle can be directly applied to the energy part. One can also view it from the minimization of energy point-of-view, as soon as the system approaches the minimum energy state, it becomes stationary). Then, the sum of specific energies inside the integral shall be the (negative) Lagrangian. The energy shall include both the potential energy due to gravitational pull and the elastic energy due to compression (later on, the terms describing the thermal energy and the rotational energy can be added): 

\begin{equation}
E_{\text{total}} = E_{\text{elastic}}+E_{\text{grav}} = \int_{\text{m=0}}^{\text{m=M}} [u-\frac{Gm}{r}] dm
\label{eq:1}
\end{equation}

where $u$ is the specific elastic energy due to compression. It should be stationary with the appropriate functional dependence of $\omega$ on $m$. Comparing it to the familiar definition of action in classical mechanics $S=\int_{\text{t=t1}}^{\text{t=t2}} L(t;x,\dot{x})  dt$, the Lagrangian of planetary interior can be identified as (negative sign is introduced for convinience): 

\begin{equation}
L(m;\omega, \dot{\omega}) = -(u-\frac{Gm}{r}) = -(u(\dot{\omega})-\frac{Gm}{(\frac{3}{4\pi})^{\frac{1}{3}} \cdot {\omega}^{\frac{1}{3}}})
\label{eq:2}
\end{equation}

Applying the Euler-Lagrange Equation to Eq.~\ref{eq:2}: 

\begin{equation}
\frac{\partial L}{\partial \omega}-\frac{d}{dm}\left( \frac{\partial L}{\partial \dot{\omega}} \right)=0
\label{eq:3}
\end{equation}

Then we have: 

\begin{equation}
\boxed{ \left(\frac{4\pi}{81}\right)^{\frac{1}{3}} G \cdot \frac{m}{{\omega}^{\frac{4}{3}}}= u''(\dot{\omega}) \cdot \ddot{\omega}}
\label{eq:main}
\end{equation}

This single 2nd-order differential equation is equivalent to the two 1st-order differential equations (mass conservation and pressure balance) that are usually used to calculate planetary interiors, just as the Euler-Lagrange Equation is equivalent to the Hamilton Canonical Equations. It is solved with the EOS (functional dependence of u on $\dot{\omega}$, $u''(\dot{\omega})=\frac{d}{d\dot{\omega}} \cdot \left( \frac{d}{d\dot{\omega}} u(\dot{\omega}) \right)$) and the following boundary conditions: 

$\begin{cases} 
\omega(0)=0, \mbox{volume is zero at the center} \\ 
\dot{\omega}(M)=v_0=\frac{1}{\rho_0}, \mbox{density is uncompressed at the surface, since there is no pressure}
\label{eq:BC}
\end{cases}$

Eq.~\ref{eq:main} can be cast into variables that people are more familiar with:

\begin{equation}
\frac{G m}{4\pi r^4}=-P'(v) \cdot \frac{dv}{dm} = -\frac{dP(v)}{dv} \cdot \frac{dv}{dm} = -\frac{dP(v)}{dm} =-\frac{dP(\frac{1}{\rho})}{dm}
\label{eq:mainfamiliar}
\end{equation}

It is no more than the pressure-balance equation, written in variable $m$ instead of $r$. 

\subsection{Implementation of EOS}

In principle, EOS expressed as the functional dependence of specific internal energy $u(v)$ on specific volume $v$ could assume any general functional form. Pressure is related as: 

\begin{equation}
P(v) = P(\frac{1}{\rho}) = -u'(v)
\end{equation}

Because $P=0$ at the surface of planet, $u'(v_0)=0$ always 

The bulk modulus $K$ is: 
\begin{equation}
K \equiv -\frac{dP}{dlnv} = -\frac{vdP}{dv} = v \cdot \frac{du'(v)}{dv} = v \cdot u''(v)  
\end{equation}

The relation between the bulk modulus at zero pressure ($K_0$), the specific internal energy $u(v)$ and the specific volume at zero pressure ($v_0=\frac{1}{\rho_0}$) is thus: 

\begin{equation}
K_0=v_0 \cdot u''(v_0) \text{ or } u''(v_0)=\frac{K_0}{v_0}={\rho}_0 K_0
\end{equation}

Many material EOSs used in Earth sciences and astrophysics are parametrized by $K_0$ (bulk modulus 
at zero-pressure) and $\rho_0$ (uncompressed density), since they are readily determined by laboratory experiments. It will be convenient to non-dimensionalize Eq~\ref{eq:main} with respect to them, so the solutions can be scaled with different $K_0$ and $\rho_0$. This is particularly useful for the power-law EOS and Birch-Murnaghan EOS to be discussed in upcoming sections. 

\subsection{Non-dimensionalization of the General Equation}

Assume the EOS can be expressed in the following form: 

\begin{equation}
P=-\frac{K_0}{\eta'(1)} \cdot \eta(f)
\label{eq:EOS}
\end{equation}

where $\eta$ is a function of $f(=\frac{\rho_0}{\rho}=\frac{v}{v_0})$ which is the fractional compression ($f \leq 1$). $\eta(1)=0$. With the substitutions of variables as: $\begin{cases} 
x \equiv \frac{m}{M}, & x \in [0,1] \\ 
y(x) \equiv \frac{\omega}{M \cdot v_0}, & y\in[0, \left( \frac{\frac{4\pi}{3}R^3}{M v_0}\right)]
\end{cases}$, 

so $y$ is differentiated with respect to $x$, $\dot{y}(x)=\frac{dy}{dx}=\frac{\dot{\omega}}{v_0}=f$, and $u''(v)=K_0 \cdot \rho_0 \cdot \frac{\eta'(f)}{\eta'(1)}$, 

Eq.~\ref{eq:main} then transforms to the following dimensionless form: 

\begin{equation}
C \cdot x= \frac{\eta'(\dot{y})}{\eta'(1)} \cdot \ddot{y} \cdot y^{\frac{4}{3}}
\label{eq:main2}
\end{equation}

where dimensionless constant $C$ is defined as: 

\begin{equation}
C \equiv \left(\frac{4\pi}{81} \right)^\frac{1}{3} \frac{G \cdot M^{\frac{2}{3}} \cdot \rho_0^{\frac{4}{3}}}{K_0} 
\label{eq:defC}
\end{equation}

Dimensionless number $C$ will later shown to be very important, as it dictates the regimes of solutions one would get, just as the dimensionless Reynolds number $Re$ does for the non-dimensionalization and scaling of the Navier-Stokes Equation in fluid dynamics. So it tells us how to scale from one solution properly to get the solutions of many other similar cases, without solving each case separately. This "self-similar" solution approach will be explored extensively when we apply Eq.~\ref{eq:main} or Eq.~\ref{eq:main2} to power-law equations of states (EOS) in the next section. 

Then, the non-dimensionalized boundary conditions become: $\begin{cases} 
y(0)=0 \\ 
\dot{y}(1)=1
\label{eq:BC2}\end{cases}$. 

Eq.~\ref{eq:main2} can be solved by shooting method: first guess an initial value of $y(1)$, then integrate inward to find $y(0)$, if $y(0)\neq$0, adjust the initial guess of $y(1)$ and iterate. 

Given the same $\eta(f)$, solution $y(x)$ only depends on $C$. So $y_1 \equiv y(1) =\frac{\frac{4\pi}{3}R^3\rho_0}{M}$ only depends on $C$ also. Define this dependence as: $y_1(C)$. Since radius $R=\left[ \frac{3M}{4\pi \rho_0} y_1(C) \right]^\frac{1}{3}$, if $y_1(C)$ can be calculated or estimated, it gives the mass-radius relation and can derive the propagation of perturbations in $K_0$ or $\rho_0$ onto mass-radius relation. $y_1(C)$ should behave as:
\begin{itemize}
\item When $C\rightarrow 0$, no compression, $y_1\rightarrow 1$, so $y_1(0)=1$. \\
\item When $C\rightarrow \infty$, infinite compression, $y_1\rightarrow 0$, so $y_1(\infty)=0$. \\
\item If $\eta(f)$ is smooth (well-behaved), $y_1(C)$ should be smooth also. 
\end{itemize}

For large $C$ ($C\gtrsim10$), solution $y(x)$ will become self-similar as one of the boundary conditions can be loosened ($\dot{y}(1)=1$ can be loosened to $\dot{y}(1)\sim1$, because the not-so-much-compressed surface layer is thin enough compared to the much-compressed bulk planet). This fact is especially useful for massive planets. 

\section{Simple Power-law EOS}

Simple power-law EOS has the following form: 

\begin{equation}
P=\frac{K_0}{k+1} \cdot \left[ \left( \frac{\rho}{\rho_0}\right)^{k+1}-1 \right]
\label{eq:powerlawEOS}
\end{equation}

equivalently, $\eta(f)=f^{-(k+1)}-1$. So $\eta'(f)=-(k+1) \cdot f^{-(k+2)}$ and $\eta'(1)=-(k+1)$. 

It is similar to the polytropic EOS in Lane-Emden equation where $P \propto \rho^{(1+\frac{1}{n})}$, where the polytropic index $n=\frac{1}{k}$. 
Eq.~\ref{eq:main2} then becomes: 

\begin{equation}
C \cdot x= {\dot{y}}^{-(k+2)} \cdot \ddot{y} \cdot y^{\frac{4}{3}}
\label{eq:main3}
\end{equation}

\subsection{Self-similar solutions}

For large $C$ ($C\gtrsim10$), if solution $y_0(x)$ for $C=C_0$ is known, solution $y(x)$ for any $C$ can be found because the solutions are self-similar. Define ratio $\lambda \equiv \frac{y}{y_0}$ and plug into Eq.~\ref{eq:main3}: 

\begin{equation}
C \cdot \lambda^{k-\frac{1}{3}} \cdot x= C_0 \cdot x
\label{eq:ss}
\end{equation}

Therefore,

\begin{equation}
\lambda=\frac{y}{y_0}=\left( \frac{C}{C_0} \right)^{\frac{1}{\frac{1}{3}-k}}
\label{eq:ss2}
\end{equation}

Recall that $C = \left(\frac{4\pi}{81} \right)^\frac{1}{3} \frac{G \cdot M^{\frac{2}{3}} \cdot \rho_0^{\frac{4}{3}}}{K_0}$, and $R=\left( \frac{3M \cdot y(1)}{4\pi \rho_0} \right)^{\frac{1}{3}}$, we have: 


\begin{equation}
\frac{R}{R_0}=\left( \frac{M}{M_0} \right)^{\frac{1}{3}} \cdot \left( \frac{y(1)}{y_0(1)} \right)^{\frac{1}{3}} \cdot \left( \frac{\rho_0}{\rho_{00}} \right)^{-\frac{1}{3}} =\left( \frac{M}{M_0} \right)^{\frac{1-k}{1-3k}} \cdot \left( \frac{\rho_0}{\rho_{00}} \right)^{\frac{1+k}{1-3k}} \cdot \left( \frac{K_0}{K_{00}} \right)^{\frac{1}{3k-1}}
\label{eq:ss3}
\end{equation}

$\rho_{00}$ and $K_{00}$ are those of $C_0$. If $\rho_0$ and $K_0$ are held the same, then the mass-radius relation for large compression is: 

\begin{equation}
R \propto M^{\frac{1-k}{1-3k}}
\label{eq:ss4}
\end{equation}

Eq.~\ref{eq:ss4} is useful to show the general behaviors of solutions of different $k$-values in next section. 

\subsection{Discussion of different $k$-values}

The value of $k$ depends on the physics governing the interior of that object: 

\begin{center}
    \begin{tabular}{ | c | c | c | c | p{9cm} |}
    \hline
    $k$ & $k+1$ & $\gamma=\frac{k}{2}+\frac{1}{3}$ & $n=\frac{1}{k}$ & Physical Scenarios \\ \hline
    
    $\frac{1}{3}$ &  $\frac{4}{3}$ & $\frac{1}{2}$ & $3$ & (1) Eddington Stellar Model~\citep{Eddington:1926} (2) extreme relativistic degenerate $e^{-}$-gas~\citep{Eliezer:2002} \\ \hline
    
   $\frac{2}{3}$ &  $\frac{5}{3}$ & $\frac{2}{3}$ & $1.5$ & (1) Uranus and Neptune (monatomic ideal gas, applicable to $10^{-6} \sim 10^1$ Mbar, obtained by fitting to EOS in~\citet{Helled:2011}) (2) non-relativistic degenerate $e^{-}$-gas~\citep{Salpeter:1967} \\ \hline
   
   $1$ &  $2$ & $\frac{5}{6}$ & $1$ & (1) Jupiter and Saturn (fluid metallic hydrogen, applicable to $10^{-3} \sim 10^2$ Mbar, obtained by fitting to EOS in~\citet{Guillot:2004}) \\ \hline
   
   $\frac{4}{3}$ &  $\frac{7}{3}$ & $1$ & $0.75$ & (1) high-pressure limit of BM2 EOS~\citep{Birch:1947} \\
   
    \hline
    \end{tabular}
\end{center}

When $k=\frac{1}{3}$, the denominator $1-3k=0$, indicating there is a critical $C_{\text{crit}}$ beyond which no solution exists. Numerically solving Eq.~\ref{eq:main3} shows that $C_{\text{crit}}\approx 1.1$ and with the appropriate $\rho_0$ and $K_0$ gives the Chandrasekhar mass limit. 

When $\frac{1}{3}<k<1$, $\frac{1-k}{1-3k}<0$, for large $C$, radius decreases with increasing mass. This is the case for white dwarfs, and also applicable to Uranus and Neptune (Neptune being more massive but slightly smaller in radius). 

When $k=1$, the numerator $1-k=0$, thus, for large $C$, radius remains constant independent of mass. This is applicable to Jupiters, super-Jupiters, and brown dwarfs, all of which have nearly identical radii. For large $C$, $y_1(C) \approx \left( \frac{\pi}{3 \sqrt{C}} \right)^3 $ so it can be shown that this radius $R \approx \left( \frac{\pi K_0}{4G{\rho_0}^2} \right)^{\frac{1}{2}}$. With $\frac{K_0}{{\rho_0}^2} \approx 4 \text{bar}/\left(\text{kg/m}^3 \right)^2$ it gives $R \approx 1R_{\text{Jupiter}} \approx 10R_{\oplus}$. 

When $k>1$, $\frac{1-k}{1-3k}>0$, radius increases with increasing mass. $k=\frac{4}{3}$ is of particular interest as it is the high-pressure limit of Birch-Murnaghan 2nd-order (BM2) EOS applicable to both iron-alloys and silicates in rocky planet interiors. BM2 will be discussed extensively in the next section. 

When $k\rightarrow+\infty$, $\frac{1-k}{1-3k}\rightarrow\frac{1}{3}$. This material has infinite rigidity meaning constant density. Therefore, $R \propto M^{\frac{1}{3}}$ is expected to be the case. 

\section{Towards a Universal Mass-Radius Relation}

\subsection{Generalized Power-Law EOS}
A generalized power-law EOS bears the following form: 

\begin{equation}
P=\frac{K_0}{k_2-k_1} \cdot \left[\left(\frac{\rho}{\rho_0}\right)^{k_2+1}-\left(\frac{\rho}{\rho_0}\right)^{k_1+1} \right]
\label{eq:EOSgeneral}
\end{equation}

where $k_2>k_1\geqslant -1$. Equivalently, we have $\eta(f)=f^{-(k_2+1)}-f^{-(k_1+1)}$. So $\eta'(f)=-(k_2+1) \cdot f^{-(k_2+2)}+(k_1+1) \cdot f^{-(k_1+2)}$, and $\eta'(1)=-(k_2-k_1)$. When $k_1=-1$, it is reduced back to the simple power-law EOS. 

This form of EOS includes the Birch-Murnaghan 2nd-order (BM2) EOS~\citep{Zeng:2016, Birch:1952,  Birch:1947} which is good for approximating the compression of iron-alloys and silicates in rocky planetary interiors, as well as the Lennard-Jones potential~\citep{Jones:1924} approximating the interaction among neutral atoms or molecules. We expect $y_1(C)$ of the generalized power-law EOS can be very well approximated by the following functional form for a certain range of $C$: 

\begin{equation}
y_1(C) \approx \frac{1}{1+ \alpha \cdot C^{\beta}}
\label{eq:y1general}
\end{equation}

where $\alpha$ and $\beta$ are constants selected based on the exact form of EOS, i.e., $k_1$ and $k_2$. Recall the definition of $y_1(C) =  \left(\frac{\frac{4\pi}{3}R^3}{M} \cdot \rho_0\right)$ or $R=\left( \frac{3M}{4\pi \rho_0} y_1(C) \right)^\frac{1}{3} $, then the general form of Mass-Radius relation for this type of EOS can be expressed as: 

\begin{equation}
\boxed{ \frac{R}{R_{\oplus}} \approx \left(\frac{a_1 \cdot (M/M_{\oplus})}{1+a_2 \cdot (M/M_{\oplus})^{a_3}} \right)^\frac{1}{3} }
\label{eq:MRgeneral}
\end{equation}

where $a_1=\frac{\rho_{\oplus}}{\rho_0}$, $a_2=\alpha \cdot \left[ \left(\frac{4\pi}{81} \right)^\frac{1}{3} \frac{G \cdot M_{\oplus}^{\frac{2}{3}} \cdot \rho_0^{\frac{4}{3}}}{K_0} \right]^{\beta}$, and $a_3 = \frac{2}{3} \beta$ are constants depending on the exact form of EOS. $M_{\oplus}=5.9724 \cdot 10^{24}$ kg and $\rho_{\oplus}=5.515$ g/cc are the mass and mean density for Earth. 

\subsection{Birch-Murnaghan EOS and Application to Rocky Planets}

Birch-Murnaghan 2nd-order (BM2) EOS provides a decent fit to material compression in rocky planetary interior of both core (good up to 12 TPa) and mantle (good up to 3.5 TPa)~\citep{Zeng:2016, Birch:1952,  Birch:1947}. These pressures approximately correspond to the central pressure and core-mantle boundary pressure of the interior of a $\sim30 \text{M}_\oplus$ rocky planet of core mass fraction (CMF)$\approx0.3$ respectively. 

BM2 EOS ($k_2=\frac{4}{3}$ and $k_1=\frac{1}{3}$) has the following form: 

\begin{equation}
P=\frac{3}{2} \cdot K_0 \left[\left(\frac{\rho}{\rho_0}\right)^{\frac{7}{3}}-\left(\frac{\rho}{\rho_0}\right)^{\frac{5}{3}} \right]=\frac{3}{2} \cdot K_0 \left[\left(\frac{v}{v_0}\right)^{-\frac{7}{3}}-\left(\frac{v}{v_0}\right)^{-\frac{5}{3}} \right]
\label{eq:BM2}
\end{equation}

The fit of BM2 to Earth's seismic density profile PREM~\citep{Dziewonski:1981} gives the following~\citep{Zeng:2016}: for lower mantle, $\rho_0=3.98$ g/cc, $K_0=206$ GPa, error$\sim1\%$ in density; for outer core, $\rho_0=7.05$ g/cc, $K_0=201$ GPa, error$\sim1\%$ in density. 

The fact that $K_0\approx200$ GPa for both lower mantle and outer core is convenient for modeling purpose. It suggests that, at any pressure, the density contrast between core and mantle remains approximately the same, including the core-mantle boundary (CMB). 

Eq.~\ref{eq:main2} then becomes: 

\begin{equation}
C \cdot x= \left({\frac{7}{2} \cdot \dot{y}}^{-\frac{10}{3}}-{\frac{5}{2} \cdot \dot{y}}^{-\frac{8}{3}} \right) \cdot \ddot{y} \cdot y^{\frac{4}{3}}
\label{eq:main4}
\end{equation}

$y_1(C)$ is solved numerically and then fit to an analytic function of C (with $<1\%$ error): 

\begin{equation}
y_1(C)\approx\frac{1}{1+0.5 \cdot C^{0.885}},~~\text{for}~0\leq C \lesssim 9
\label{eq:y1_2}
\end{equation}

Therefore, the general mass-radius relation for BM2 EOS is: 

\begin{equation}
\boxed{ \frac{R}{R_{\oplus}} \approx \left(\frac{(M/M_{\oplus})}{(\rho_0/\rho_{\oplus})} \cdot \frac{1}{1+0.306 \cdot \left[ \frac{(\rho_0/{\rho_{\oplus}})^2}{(K_0/200\text{GPa})^{3/2}} \right]^{0.59}  \cdot (M/M_{\oplus})^{0.59}} \right)^\frac{1}{3} }
\label{eq:MR}
\end{equation}

Strictly speaking, Eq.~\ref{eq:MR} only applies to one-layer planet. However, since $K_0 \approx$ 200 GPa for both core and mantle, it can be used for two-layer rocky planets with the equivalent uncompressed average density $\boxed{\rho_0 = \left(3.86+2 \cdot \text{CMF} + \text{CMF}^3 \right) \text{g/cc}}$. It is applicable to $0.3 \sim 30$ M$_{\oplus}$ with fractional error in radius generally less than 1$\%$. 

\subsection{Propagation of EOS uncertainties onto mass and radius}

With Eq.~\ref{eq:MR}, one can estimate the propagation of EOS uncertainties (in both $\rho_0$ and $K_0$, which are usually experimentally determined) onto mass and radius. For large mass, we could neglect the 1 in the denominator to get: 

\begin{equation}
\frac{R}{R_{\oplus}} \sim \left( \frac{M}{M_{\oplus}} \right)^{0.137} \cdot \left( \frac{\rho}{\rho_{\oplus}} \right)^{-0.727} \cdot \left( \frac{K_0}{200\text{GPa}} \right)^{0.295}
\label{eq:MRsim}
\end{equation}

Taking natural logarithm of Eq.~\ref{eq:MRsim} on each side and differentiate, we get: 

\begin{equation}
\frac{\delta R}{R} \approx 0.137 \cdot \frac{\delta M}{M}-0.727 \cdot \frac{\delta \rho_0}{\rho_0}+0.295 \cdot \frac{\delta K_0}{K_0}
\label{eq:RdR}
\end{equation}

Therefore, the perturbation effect of $K_0$ is about one-third that of $\rho_0$, which is slightly less than unity. As expected, an increase in density will make the planet smaller, while an increase in bulk modulus will make the planet bigger. 

\section{Thermal Effect}

\subsection{Adiabatic Temperature Profile}

It is generally attested that throughout most of planetary interiors, except the boundary layers, the temperature gradient is near adiabatic due to convection that preserves specific entropy. For an adiabatic Debye solid, the temperature $T$ and density $\rho$ are related by: 

\begin{equation}
T \propto \rho^{\gamma}
\label{eq:TE1}
\end{equation}

Here $\gamma$ is the Gr\"{u}neisen parameter for solid, not to be confused with the adiabatic index for gas, since planets with solid interiors are mostly concerned with here. It can be shown that the following relation holds for any Debye solid~\citep{Vocadlo:1994, Slater:1939}: 

\begin{equation}
\gamma=-\frac{1}{6}+\frac{1}{2} \cdot \frac{d\ln K}{d\ln \rho} = -\frac{1}{6}+\frac{1}{2} \cdot \frac{dK}{dP}=-\frac{1}{6}+\frac{1}{2} \cdot K'
\label{eq:TE2}
\end{equation}

In particular, for the simple power-law EOS of $P \sim \left(\rho^{k+1}+\text{const}\right)$, $K'=k+1$, so we have the following simple relation between $\gamma$ and $k$: 

\begin{equation}
\gamma=\frac{k}{2}+\frac{1}{3}
\label{eq:gammak}
\end{equation}

and vice versa, 

\begin{equation}
k=2 \cdot \left (\gamma-\frac{1}{3} \right)
\label{eq:kgamma}
\end{equation}

So if $\gamma \approx$const within a certain range of pressure, then $K' \approx$const and $k \approx$const within that range as well, then $P$ must have a power-law dependence on $\rho$ with power-index $k$ in that range, and vice versa. 

The thermal energy is mostly contributed by translational vibration of atoms in their crystal lattices, while electron contribution is small because of being degenerate. Above Debye temperature $\theta_D$, usually true for planetary interiors, the molar heat capacity of any solid is $\sim3R$ due to $3$ translational modes of vibration. 
Debye theory shows that the specific thermal energy can be expressed as: 

\begin{equation}
u_{th}(\dot{\omega},\sigma) =\frac{3RT}{\mu} =\frac{3R\theta_{D0}}{\mu} \cdot \exp \left( \frac{\mu s}{3R} \right) \cdot (\rho_0 \dot{\omega})^{-\gamma}  \propto {\dot{\omega}}^{-\gamma}
\label{eq:TE3}
\end{equation}

where $\theta_{D0}$ is the Debye temperature of this solid under no compression, $\mu$ is the average atomic weight of the mineral, and $s \equiv \frac{3R}{\mu} \ln\left( \frac{T}{\theta_D} \right)$ is the specific entropy at temperature T if the specific entropy at $\theta_D$ is assumed to be 0.

\subsection{Melting Temperature Profile}

$T_{\text{melting}}$ (melting-temperature) profile generally has a different slope from that of the adiabat. 
Lindemann criterion~\citep{Lindemann:1910} describes the melting of solids as lattice vibrational amplitude exceeds a certain threshold of the lattice spacing. Combining it with the Debye theory gives: 

\begin{equation}
f_{\text{melting}} \equiv \frac{\text{lattice vibration amplitude}}{\text{lattice spacing}} = \frac{{\langle u^2 \rangle}^{\frac{1}{2}}}{a} = \left( \frac{R \cdot T_{\text{melting}}}{{v_{\text{seismic}}}^2 \cdot \mu} \right)^{\frac{1}{2}} \approx 0.1
\label{eq:Melting1}
\end{equation}

where ${\langle u^2 \rangle}^{\frac{1}{2}}$ is the root-mean-square displacement of an atom, $a$ is the lattice spacing, and $v_{\text{seismic}} \sim \theta_{D} \cdot \left( \frac{M_{\text{atom}}}{\rho} \right)^{\frac{1}{3}} $ is the bulk seismic velocity (mean sound speed). This gives: 

\begin{equation}
T_{\text{melting}} \propto {\theta_{D}}^2 \cdot {\rho}^{-\frac{2}{3}} \propto {\rho}^{2 \cdot \left (\gamma-\frac{1}{3} \right)} = {\rho}^{k} \propto {\dot{\omega}}^{-k}
\label{eq:Melting2}
\end{equation}

$k$ is the index of power-law EOS defined earlier. Therefore, the slopes are $\begin{cases} 
\frac{d\ln T_{\text{melting}}}{d\ln \rho}=k \\ 
\frac{d\ln T_{\text{adiabat}}}{d\ln \rho}=\gamma
\label{eq:Melting3}\end{cases}$. 

When $k=\gamma=\frac{2}{3}$, the two slopes are equal. Generally, $k>\frac{2}{3}$ for solid planet interior, so $k>\gamma$ (melting curve is steeper than adiabat). As a result, melting always occurs near the top within a uniform region inside a solid planet. This explains why Earth's inner core is solid while outer core is liquid (the inner-outer core boundary is where the melting curve intersects the adiabat) but not the other way around. This also explains why Earth's uppermost part of the entire mantle is most susceptible to partial melting. When a magma ocean was present early on, it must be at the surface also due to this reason. Concordantly, if the heat content of a convective solid planet is increased somehow, the planet will melt from top downward. On the other hand, if the planet cools gradually, it will freeze from center outward.

\section{Rotational Effect}

The total angular momentum $J$ of a planet can be expressed as the product of its moment of inertia $I$ and its rotational angular frequency $\Omega \equiv \frac{2\pi}{\text{Period}}$: 

\begin{equation}
J=I \cdot \Omega
\label{eq:rot01}
\end{equation}

The total rotational kinetic energy $E_{\text{rot}}$ is: 

\begin{equation}
E_{\text{rot}} = \frac{1}{2} \cdot I \cdot {\Omega}^2 = \frac{J^2}{2\cdot I}
\label{eq:rot02}
\end{equation}

A small variation of $E_{\text{rot}}$ in consideration of $J$ being conserved is: 

\begin{equation}
\delta E_{\text{rot}} = \delta \left( \frac{J^2}{2\cdot I} \right) = -\frac{J^2}{2\cdot I^2} \cdot \delta I
\label{eq:rot02}
\end{equation}

Assuming the planet is not spinning too fast to be significantly distorted from a spherical shape, the momentum of inertia about the rotational axis can be calculated as: 

\begin{equation} 
I = \iiint_V (x^2 + y^2) dm \approx \frac{2}{3} \iiint_V r^2 dm = \frac{2}{3} \cdot \left( \frac{3}{4\pi} \right)^{2/3} \cdot \int_{\text{m=0}}^{\text{m=M}} \omega^{2/3} dm
\label{eq:rot03}
\end{equation}

Thus, the specific rotational energy goes like: 

\begin{equation} 
u_{\text{rot}} \sim \omega^{-2/3}
\label{eq:rot04}
\end{equation}

\section{Virial Theorem}

Euler-Lagrange Equation (Eq.~\ref{eq:3}) gives 

\begin{equation}
\frac{\partial L}{\partial \omega}=\frac{d}{dm}\left( \frac{\partial L}{\partial \dot{\omega}} \right)
\label{eq:virial01}
\end{equation}

Multiply both sides by $\omega$ and integrate from 0 to M, 

\begin{equation}
\int_{\text{m=0}}^{\text{m=M}} \omega \cdot \frac{\partial L}{\partial \omega} \cdot dm = \int_{\text{m=0}}^{\text{m=M}} \omega \cdot \frac{d}{dm}\left( \frac{\partial L}{\partial \dot{\omega}} \right) \cdot dm
\label{eq:virial02}
\end{equation}

The RHS can be integrated by parts as: 

\begin{equation}
\int_{\text{m=0}}^{\text{m=M}} \omega \cdot d\left( \frac{\partial L}{\partial \dot{\omega}} \right) = \omega \cdot \left( \frac{\partial L}{\partial \dot{\omega}} \right) \bigg|_{\text{m=0}}^{\text{m=M}} - \int_{\text{m=0}}^{\text{m=M}} d\omega \cdot \left(  \frac{\partial L}{\partial \dot{\omega}} \right)
\label{eq:virial03}
\end{equation}

At m=0, $\omega=0$. At m=M, $\frac{\partial L}{\partial \dot{\omega}}=0$. Therefore, the term $\omega \cdot \left( \frac{\partial L}{\partial \dot{\omega}} \right) \bigg|_{\text{m=0}}^{\text{m=M}}$ vanishes. 

Then, we have: 

\begin{equation}\label{eq:virial04}
\int_{\text{m=0}}^{\text{m=M}} \frac{\partial L}{\partial \ln \omega} \cdot dm = -\int_{\text{m=0}}^{\text{m=M}} \frac{\partial L}{\partial \dot{\omega}} \cdot d\omega = -\int_{\text{m=0}}^{\text{m=M}} \frac{\partial L}{\partial \dot{\omega}} \cdot \dot{\omega} \cdot dm = -\int_{\text{m=0}}^{\text{m=M}} \frac{\partial L}{\partial \ln \dot{\omega}} \cdot dm
\end{equation}

Collecting terms to one side, we thus obtain the following form of virial theorem: 

\begin{equation}
\boxed{ \int_{\text{m=0}}^{\text{m=M}} \left[  \frac{\partial}{\partial \ln{\omega}}+\frac{\partial}{\partial \ln{\dot{\omega}}}\right] L(m;\omega, \dot{\omega}) \cdot dm = 0 }
\label{eq:virial1}
\end{equation}
 
This result can also be viewed from the variational principle itself, by considering a small variation of the total action $S$ about the equilibrium: 

\begin{equation}\label{eq:virial05}
\delta S = \int_{\text{m=0}}^{\text{m=M}} \left( \frac{\partial L}{\partial \omega} \delta\omega + \frac{\partial L}{\partial \dot{\omega}} \delta\dot{\omega} \right) dm = 0
\end{equation}

If we pick a particular small variation as $\delta \omega = \alpha \cdot \omega$ where $\alpha$ is a small number (constant), then $\delta \dot{\omega} = \alpha \cdot \dot{\omega}$. 
It satisfies the one of the boundary conditions at m=0 automatically. However, it seems to violate the other boundary condition of $\dot{\omega}(\text{M})=\frac{1}{\rho_0}$ at m=M with this proportional variation. But noticing that the pressure $p=\frac{\partial L}{\partial \dot{\omega}}$ is zero at the surface, so the effect of this variation vanishes at the surface (m=M) also. Therefore, by adopting this particular choice of $\delta \omega$, the same conclusion is reached:  

\begin{equation}\label{eq:virial06}
\delta S = \int_{\text{m=0}}^{\text{m=M}} \left( \frac{\partial L}{\partial \omega} \alpha \cdot \omega + \frac{\partial L}{\partial \dot{\omega}} \alpha \cdot \dot{\omega} \right) dm = \alpha \cdot \int_{\text{m=0}}^{\text{m=M}} \left( \frac{\partial L}{\partial \ln\omega}+ \frac{\partial L}{\partial \ln\dot{\omega}} \right) dm = 0
\end{equation}

From this perspective, virial theorem can be understood as a special case or a direct consequence of the variational principle (stationary action principle) itself. 
 
Recall the definition of $L(m;\omega, \dot{\omega})$, which is the negative of the sum of specific energies, including $u_{\text{grav}}(m;\omega)$ (specific gravitational potential energy), $u_e(\dot{\omega})$ (specific internal energy due to compression, mostly contributed by electron degeneracy thus the notation), $u_{\text{th}}(\dot{\omega},s)$ (specific thermal energy due to temperature, that is, the vibrational motion of atoms in crystal lattices), and $u_{\text{rot}}$ (specific rotational kinetic energy): 
 
 \begin{equation}
L=-\left( u_{\text{grav}} + u_e + u_{\text{th}} + u_{\text{rot}} \right)
\label{eq:virial2}
\end{equation}

Each term has a different power-law dependence on ${\omega}$ or $\dot{\omega}$: 

$\begin{cases} 
u_{\text{grav}} \propto {\omega}^{-\frac{1}{3}}, \mbox{gravitational potential is inversely proportional to radius} \\ 
u_e \propto {\dot{\omega}}^{-k}, \mbox{ $k \equiv \frac{d\ln u_e}{d\ln \dot{\omega}}$ is the instantaneous power-index which can be variable location-wise} \\
u_{\text{th}} \propto {\dot{\omega}}^{-\gamma}, \mbox{ $\gamma \equiv \frac{d\ln u_{\text{th}}}{d\ln \dot{\omega}}$ is the instantaneous power-index (Gr\"{u}neisen parameter)} \\
u_{\text{rot}} \propto {\omega}^{-2/3}, \mbox{specific rotational energy} 
\label{eq:virial3}
\end{cases}$

All the terms that have to do with $\omega$ are long-range global interactions, due to gravitational pull or rotation, and when viewed from the point of general relativity, are due to the distortion of space-time fabrics.
All the terms that have to do with $\dot{\omega}$ are short-range local interactions, due to the thermal motions or quantum interactions among electrons and atoms, and when viewed from the point of quantum physics, are due to entropy in particular. 

Eq.~\ref{eq:virial1} then becomes: 

\begin{equation}
\int_{\text{m=0}}^{\text{m=M}} \left(\frac{1}{3} \cdot u_{\text{grav}} +  k \cdot u_e +  \gamma \cdot u_{\text{th}} + \frac{2}{3} \cdot u_{\text{rot}} \right) \cdot dm= 0
\label{eq:virial4}
\end{equation}

$u_{\text{grav}}<0$, $u_e>0$, $u_{\text{th}}>0$, $u_{\text{rot}}>0$, equivalently, we have: 
\begin{equation}
\boxed{ \frac{1}{3} \cdot {\text{E}}_{\text{grav}} + \overline{k} \cdot {\text{E}}_e + \overline{\gamma} \cdot {\text{E}}_{\text{th}} +  \frac{2}{3} \cdot E_{\text{rot}} = 0 }
\label{eq:virial5}
\end{equation}
$\overline{k}$ and $\overline{\gamma}$ imply the average over the integral. 
The Eq.~\ref{eq:virial5} suggests that energy could be exchanged in between all these terms during the secular evolution of a planet, while always satisfying this identity. 
And $E_{\text{grav}}<0$, $E_e>0$, $E_{\text{th}}>0$, $E_{\text{rot}}>0$. It is easy to know the "$+$" and "$-$" sign of each term. Have the following thought experiment: think about a planet contracts slightly, its gravitational energy becomes more negative, while its compression thus degeneracy energy shall increase, and due to adiabatic compression its thermal energy shall increase, and due to the conservation of angular momentum, its rotational kinetic energy shall increase also. 

The total energy of the system ${\text{E}}_{\text{tot}}={\text{E}}_{\text{grav}}+{\text{E}}_e+{\text{E}}_{\text{th}}+E_{\text{rot}}$. And since planet is a bound system, we expect ${\text{E}}_{\text{tot}}<0$. 

Usually, $ \mid {\text{E}}_{\text{grav}} \mid \sim \mid {\text{E}}_e \mid \gg \mid{\text{E}}_{\text{th}} \mid \gg \mid{\text{E}}_{\text{rot}} \mid$ for planetary interior. 

For Earth, 
\begin{itemize}

\item $\mid {\text{E}}_{\text{grav},\oplus} \mid = \frac{2}{3}\frac{GM_p^2}{R_p} \approx 2.5 * 10^{32}$J

\item $\mid {\text{E}}_{\text{differentiation}, \oplus} \mid \approx \frac{1}{15}\frac{GM_p^2}{R_p} \approx 2.5 * 10^{31}$J $\approx \frac{1}{10} \cdot \mid {\text{E}}_{\text{grav},\oplus} \mid$, see~\citet{Zeng:2016c} for detailed derivation. 

\item $\mid {\text{E}}_{\text{th}, \oplus} \mid \approx (\text{effective mantle heat capacity}) \cdot (\text{mantle potential temperature})$ 

$ = \left( \frac{3R}{\mu}\cdot M_p \approx 7.5 \cdot 10^{27} \text{J/K} \right) \cdot (1700\text{K}) \approx 1.3 \cdot 10^{31}$J $\approx \frac{1}{2} \cdot \mid {\text{E}}_{\text{differentiation}, \oplus} \mid$, see~\citet{Zeng:2016c} for detailed derivation. 

\item $\mid {\text{E}}_{\text{rot}, \oplus} \mid \approx 2 * 10^{29}$J $\approx \frac{1}{65} \cdot \mid {\text{E}}_{\text{th}, \oplus} \mid$. It is now small compared to other terms, but early on in Earth's history, it is much bigger and of comparable magnitute with other terms, especially after the giant impacts. 

\end{itemize}

\section{Conclusion}

Here in this paper we present a new framework of formulating the planetary interior based on the general variational (stationary action) principle. 

From this principle, a single second-order differential equation describing the planetary interior, which is equivalent to the two first-order differential equations (pressure balance and mass conservation), is derived. 
This second-order differential equation can be non-dimensionalized for simplicity with the introduction of a dimensionless constant $C \equiv \left(\frac{4\pi}{81} \right)^\frac{1}{3} \frac{G \cdot M^{\frac{2}{3}} \cdot \rho_0^{\frac{4}{3}}}{K_0} $, which characterizes the degree of compression. 

By implementing different EOSs, primarily power-law EOS with different power-index $k$, and the Birch-Murnaghan second-order (BM2) EOS, which is a modified power-law, applicable to terrestrial planet interior, different solutions are categorized and discussed. An emphasis is placed upon deriving a universal mass-radius relation for rocky planets, and the propagation of errors of EOS onto the mass-radius relation. A discussion of the thermal adiabatic temperature gradient, the melting temperature gradient, the rotational effect, and a form of viral theorem applicable to planetary interiors, are also provided. 

We hope that this paper presents a new perspective of planetary interior, as an entity requiring stationary action in both time and mass dimensions. This approach shall remain valuable for the current field of exoplanet research, as great number of planets are being measured, but with limited accuracies in mass, radius and other measurable quantities, similar to what we have encountered for the study of stellar interiors in the early 20th century.

\section{Acknowledgement}
This work was supported by a grant from the Simons Foundation (SCOL [award \#337090] to L.Z.).
Part of this research was conducted under the Sandia Z Fundamental Science Program and supported by the Department of Energy National Nuclear Security
Administration under Award Numbers DE-NA0001804 and DE-NA0002937 to S. B. Jacobsen (PI) with Harvard University. This research is the authors' views and not those of the DOE. The authors would like to thank Dimitar D. Sasselov for insightful suggestions and helpful comments on this paper. The author Li Zeng would like to thank Beatrice Chrystal and Charles Hallisey for teaching him P\={a}li, the ancient Canonical language in Therav\={a}da Buddhism, the philosophy of which on kamma (action) inspired the seminal idea of this paper.   

\clearpage


\bibliographystyle{apj}
\bibliography{mybib}

\end{document}